\documentclass[structabstract,longauth]{aa}

\usepackage{natbib,twoopt}
\usepackage[breaklinks=true]{hyperref} 
\bibpunct{(}{)}{;}{a}{}{,} 
\usepackage{graphicx}
\usepackage{amssymb}
\usepackage{longtable}
\usepackage{dpfloat}
\usepackage{subfig}
\usepackage{txfonts}
\usepackage{journal_names}

\begin{document}

\title{Galaxy And Mass Assembly (GAMA): The mass-metallicity relationship}

\author{C.~Foster\thanks{cfoster@eso.org}\inst{1,2} 
\and A.M.~Hopkins\inst{3}
\and M.~Gunawardhana\inst{4}
\and M.A. Lara-L\'opez\inst{3} 
\and R.G.~Sharp\inst{5}
\and O.~Steele\inst{6}
\and E.N.~Taylor\inst{4}
\and S.P.~Driver\inst{7,8} 
\and I.K.~Baldry\inst{9}
\and S.P.~Bamford\inst{10}
\and J.~Liske\inst{11}  
\and J.~Loveday\inst{12} 
\and P.~Norberg\inst{13} 
\and J.A.~Peacock\inst{14}
\and {M. Alpaslan}\inst{7,8}
\and {A.E.~Bauer}\inst{3}
\and {J. Bland-Hawthorn}\inst{4}  
\and S.~Brough\inst{3} 
\and E.~Cameron\inst{15} 
\and M.~Colless\inst{3} 
\and C.J.~Conselice\inst{10} 
\and S.M.~Croom\inst{4} 
\and C.S.~Frenk\inst{13} 
\and D.T.~Hill\inst{8}
\and D.H.~Jones\inst{16} 
\and L.S.~Kelvin\inst{7,8} 
\and K.~Kuijken\inst{17} 
\and R.C.~Nichol\inst{6} 
\and M.S.~Owers\inst{3}
\and H.R.~Parkinson\inst{14} 
\and K.A.~Pimbblet\inst{16} 
\and C.C.~Popescu\inst{18} 
\and M.~Prescott\inst{9} 
\and A.S.G.~Robotham\inst{7,8} 
\and A.R.~Lopez-Sanchez\inst{3,19}
\and W.J.~Sutherland\inst{20} 
\and D.~Thomas\inst{6,21} 
\and R.J.~Tuffs\inst{22}
\and E.~van~Kampen\inst{11} 
\and D.~Wijesinghe\inst{4} }

\institute{European Southern Observatory, Alonso de Cordova 3107, Vitacura, Santiago, Chile 
\and Centre for Astrophysics \& Supercomputing, Swinburne University, Hawthorn, VIC 3122, Australia 
\and Australian Astronomical Observatory, PO Box 296, Epping, NSW 1710, Australia 
\and Sydney Institute for Astronomy, School of Physics, University of Sydney, NSW 2006, Australia 
\and The Australian National University, Mount Stromlo Observatory, Cotter Road, Weston Creek, ACT, 2611, Australia 
\and Institute of Cosmology and Gravitation (ICG), University of Portsmouth, Dennis Sciama Building, Burnaby Road, Portsmouth PO1 3FX, UK 
\and International Centre for Radio Astronomy Research, The University of Western Australia, 35 Stirling Highway, Crawley, WA 6009, Australia 
\and School of Physics \& Astronomy, University of St Andrews, North Haugh, St Andrews, KY16 9SS, UK 
\and Astrophysics Research Institute, Liverpool John Moores University, Twelve Quays House, Egerton Wharf, Birkenhead, CH41 1LD, UK 
\and Centre for Astronomy and Particle Theory, University of Nottingham, University Park, Nottingham NG7 2RD, UK 
\and European Southern Observatory, Karl-Schwarzschild-Str.~2, 85748 Garching, Germany 
\and Astronomy Centre, University of Sussex, Falmer, Brighton BN1 9QH, UK 
\and Institute for Computational Cosmology, Department of Physics, Durham University, South Road, Durham DH1 3LE, UK 
\and Institute for Astronomy, University of Edinburgh, Royal Observatory, Blackford Hill, Edinburgh EH9 3HJ, UK 
\and Department of Physics, Swiss Federal Institute of Technology (ETH-Z\"urich), 8093 Z\"urich, Switzerland 
\and School of Physics, Monash University, Clayton, Victoria 3800, Australia 
\and Leiden University, P.O.~Box 9500, 2300 RA Leiden, The Netherlands 
\and Jeremiah Horrocks Institute, University of Central Lancashire, Preston PR1 2HE, UK 
\and Department of Physics and Astronomy, Macquarie Uiversity, NSW 2109, Australia 
\and Astronomy Unit, Queen Mary University London, Mile End Rd, London E1 4NS, UK 
\and South-East Physics Network (SEPnet) 
\and Max Planck Institute for Nuclear Physics (MPIK), Saupfercheckweg 1, 69117 Heidelberg, Germany 
}

\date{Accepted September 2012}

\abstract{The mass-metallicity relationship (MMR) of star-forming galaxies is well-established, however there is still some disagreement with respect to its exact shape and its possible dependence on other observables.}
{We measure the MMR in the Galaxy And Mass Assembly (GAMA) survey. We compare our measured MMR to that measured in the Sloan Digital Sky Survey (SDSS) and study the dependence of the MMR on various selection criteria to identify potential causes for disparities seen in the literature.}
{We use strong emission line ratio diagnostics to derive oxygen abundances. We then apply a range of selection criteria for the minimum signal-to-noise in various emission lines, as well as the apparent and absolute magnitude to study variations in the inferred MMR.}
{The shape and position of the MMR can differ significantly depending on the metallicity calibration and selection used. After selecting a robust metallicity calibration amongst those tested, we find that the mass-metallicity relation for redshifts $0.061\lesssim z\lesssim0.35$ in GAMA is in reasonable agreement with that found in the SDSS despite the difference in the luminosity range probed.}
{In view of the significant variations of the MMR brought about by reasonable changes in the sample selection criteria and method, we recommend that care be taken when comparing the MMR from different surveys and studies directly. We also conclude that there could be a modest level of evolution over $0.06\le z\le0.35$ within the GAMA sample.}


\titlerunning{The mass-metallicity relationship in GAMA}

\maketitle

\section{Introduction}\label{sec:introduction}
\defcitealias{Kewley02}{KD02}
\defcitealias{Kewley08}{KE08}
\defcitealias{Kobulnicky04}{KK04}
\defcitealias{LaraLopez10b}{L10b}
\defcitealias{Mannucci10}{M10}
\defcitealias{McGaugh91}{M91}
\defcitealias{Tremonti04}{T04}
\defcitealias{Pettini04}{PP04}
The mass-metallicity relationship (MMR) describing a correlation between stellar mass and gas-phase metallicity in galaxies was first reported by \citet{Lequeux79}. Looking at spectra of H\,{\sc ii} regions for a small sample of irregular and blue compact galaxies, they found that metallicity and stellar mass were correlated over two orders of magnitudes in mass. Subsequently, with the advent of large spectroscopic surveys of galaxies, the reality of the MMR has been confirmed and established many times since, for both stellar \citep[e.g.,][]{Gallazzi05,Mendel09} and gas-phase \citep[e.g.,][]{Tremonti04,Zahid11} metallicities.

Several mechanisms have been proposed to explain the origins of the MMR. These include outflows, with enriched gas being retained preferentially by high-mass galaxies due to the increased depth of their potential well \citep[e.g.,][]{Larson74b,Kobayashi07,Spitoni10}. Another possible mechanism is the interplay between the outflow of enriched gas and infall of pristine gas from the inter-galactic medium \citep[e.g.,][]{Finlator08}. The downsizing phenomenon, wherein stars found in higher-mass galaxies today formed rapidly at earlier epochs suggests the surrounding gas should be enriched early. Stars found in low mass, low redshift galaxies, however, have lower $\alpha$-element abundance ratios, suggesting they formed slowly over longer periods \citep[e.g.,][]{Cowie96,Kodama04,Abraham05,Thomas10}. Indeed, \citet{Garnett02} suggest that the MMR can be explained by the fact that low mass galaxies have higher gas fractions than high mass ones, implying that they are still converting gas into stars. Moreover, \citet{Rodrigues12} show that galaxies at redshift $z\sim2.2$ have higher gas fractions than their local counterparts, suggesting that evolution of the MMR should be expected under this scenario. A fourth possible mechanism is a stellar mass dependent or star formation rate (SFR) dependent evolving initial mass function (IMF), which has also been proposed as a viable scenario e.g., \citealt{Koppen07}; \citealt{Wilkins08a}; \citealt{Wilkins08b}; \citealt{Spitoni10}; \citealt{Gunawardhana11}; \citealt{Ferreras12}). This would also impact the rate and level of metallicity enrichment with galaxy stellar mass.

The Sloan Digital Sky Survey (SDSS) has allowed for the unequivocal confirmation of the MMR \citep[][hereafter T04]{Tremonti04}. The method used in \citetalias{Tremonti04} is based on simultaneous fitting to all available emission lines using stellar population models. While using all the available lines to make a best estimate, this approach is more sensitive to noise. This method is ultimately tied to the reliability of the models used, and artificially produces quantised output metallicities as finite values of metallicities are modeled. \citetalias{Tremonti04} provide a conversion from the popular $R23=\rm ([O\thinspace\textsc{ii}]\lambda3727+[O\thinspace\textsc{iii}]\lambda\lambda4959,5007)/H\beta$ strong emission line ratio to their calibration.

\citet[][hereafter KE08]{Kewley08} have used the SDSS data to compare the inferred metallicities using various calibrations. Empirical conversions between the different metallicity calibrations are provided. They find that the absolute position and shape of the MMR can vary significantly from one calibration to another, emphasising the need to use similar metallicity calibrations when comparing different samples. Indeed, \citetalias{Kewley08} and other authors \citep[e.g.][]{LopezSanchez10,Bresolin09,Moustakas10} have shown that metallicities derived using calibrations based on photoionization models (e.g., \citealt{McGaugh91}; \citetalias{Kewley02}; \citetalias{Kobulnicky04}) tend to be systematically 0.2-0.4~dex higher than metallicities derived using the direct method (i.e., using a direct estimation of the electron temperature) or calibrations based on it (e.g., \citetalias{Pettini04}; \citealt{Pilyugin05}). In addition, depending on the metallicity calibration used, \citetalias{Kewley08} found that the turnover of the MMR depends on the aperture covering fraction. To avoid such effects, they recommend a minimum redshift of $z=0.04$ for the SDSS $3"$ aperture fibre spectra. Similarly, \citet{Moustakas11} find a measurable dependence of the MMR on the metallicity calibration used and aperture biases using a sample of $\sim3000$ galaxies from the AGN and Galaxy Evolution Survey (AGES). They also find that the MMR is affected by contamination from Active Galactic Nuclei (AGN). Hence, differences induced by metallicity calibrations, aperture effects and AGN contamination need to be taken into account when comparing results from different surveys and studies.

Several groups have studied the redshift evolution of the MMR and sometimes find conflicting results, particularly at lower redshifts. Both \citet{LaraLopez09a,LaraLopez09b} and \citet{Savaglio05} detect evolution of the MMR up to $z=0.4$ using SDSS, and over $0.4\le z\le1.0$ from a combined sample from the Gemini Deep Deep Survey (GDDS) and the Canada-France Redshift Survey (CFRS), respectively. In sharp contrast, \citet{Carollo01} do not find significant evolution when comparing the MMR in a sample of galaxies at $0.5\le z \le0.7$ from CFRS to that locally observed. Yet, several groups independently detect significant evolution at comparable redshifts $0.4\lesssim z\lesssim1.0$ \citep[see e.g.,][]{Kobulnicky04,Mouhcine06,Rodrigues08,Morelli12}.
Pushing the redshift barrier further, \citet{Erb06} analyse a sample of 87 UV selected star forming galaxies to show that galaxies at high redshift ($z\sim 2$) not only have lower oxygen abundances, but also higher gas fraction ($\sim 50$ per cent). At redshifts $z>3$, \citet{Maiolino08} select a variety of independent ratios and calibrations to obtain metallicities and detect an evolution of the MMR and its slope in a sample of 9 star forming galaxies observed with ESO-VLT. They claim that no hierarchical formation simulation reproduces the MMR at $z\sim 3$, while monolithic collapse simulations do. At even higher redshift, \citet{Laskar11} obtain a sample of 20 gamma ray bursts at redshifts $3\le z \le 5$ and demonstrate that the MMR at high redshift is significantly offset to lower metallicities than that at $z\lesssim3$. 

Another fundamental observable parameter that has proven to be important in order to disentangle which scenario(s) is (are) responsible for the MMR is the SFR. Here again, there are some discrepancies in the various findings about the dependence of the MMR on SFR. \citet{Ellison08} select galaxies in the SDSS to study the dependence of the MMR on the specific SFR (i.e., SFR per unit mass, hereafter SSFR). They observe that galaxies with higher SSFR for their stellar mass have lower metallicities ($\log(O/H)$ is as much as 0.2 dex lower). More recently, \citet[][hereafter M10]{Mannucci10} have also used SDSS galaxies to study the inter-dependence between mass, metallicity and SFR. They claim that there exists a `fundamental relationship' or three-dimensional curved surface in stellar mass, SFR and metallicity space where all star forming galaxies lie. They also find that higher SFR galaxies (and lower metallicities) are selected at higher redshift, leading to an \emph{apparent} evolution of the MMR. \citetalias{Mannucci10} demonstrate that high redshift galaxies from \citet{Erb06}, but not \citet{Mannucci09} follow this `fundamental relationship'. Similarly, \citet[][]{LaraLopez10a} and \citet[][hereafter L10b]{LaraLopez10b} discovered a relationship which is best described as a ``fundamental plane'' between SFR, metallicity and stellar mass \citep[recently updated by][]{LaraLopez12}. In contrast to \citetalias{Mannucci10}, they do not find any curvature in three-dimensions or any evolution of the fundamental plane up to $z\sim3.5$, and detect a shallow but positive correlation between SFR and metallicity such that high SFR galaxies tend to have higher metallicities. In general, they observe that metallicities are lower, SFRs are higher and the morphology of galaxies shows a higher fraction of late-type galaxies at high redshift compared to locally. Using semi-analytic models, \citet{Yates12} reproduce the 3-dimensional surface theoretically. They explain that low SFR, high mass galaxies have typically exhausted their gas reservoirs in a recent major merger, hence preventing further star formation. Subsequent inflow of metal-poor gas then dilutes the gas around these galaxies without significant further star formation.

The focus of this paper is to present the MMR using data from the Galaxy And Mass Assembly survey \citep[GAMA, see][]{Driver11}\footnote{http://www.gama-survey.org/} and study its dependence on sample selection. A companion paper (Lara-L\'opez et al., in preparation) presents the joint distribution of GAMA and SDSS galaxies in the three-dimensional SFR/stellar mass/metallicity parameter space. GAMA is a good complement to the previous SDSS-based studies as it probes to fainter magnitudes than SDSS, in principle allowing for a range of lower stellar masses to be studied at a given redshift, and to higher redshift at a given stellar mass. The paper is structured as follows. The data and sample selection are described in Section~\ref{sec:data}. Sections~\ref{sec:analysis} and \ref{sec:results} present our analysis and highlight our results, respectively. A summary and our conclusions are given in Section~\ref{sec:concl}.

Throughout we assume a cosmology given by $\Omega_M=0.3$, $\Omega_\Lambda=0.7$ and $H_0=70$\,km\,s$^{-1}$\,Mpc$^{-1}$.

\begin{figure}
\begin{center}
\includegraphics[width=84mm]{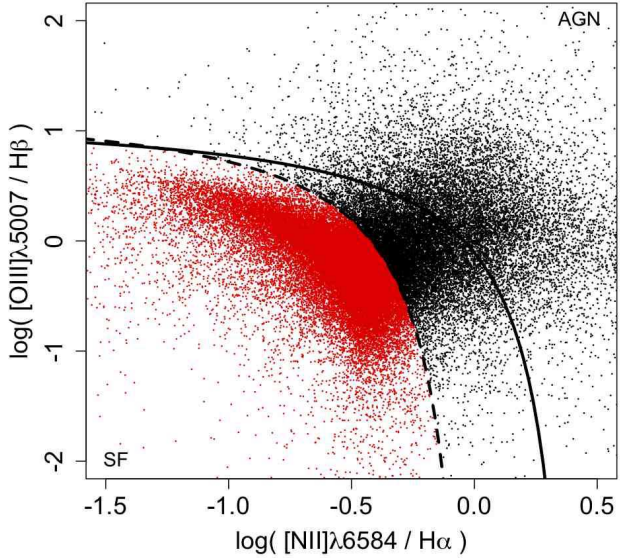}
\caption{Selection of star-forming galaxies (red) using the prescription from \citet[][dashed line]{Kauffmann03} BPT diagram \citep{Baldwin81}. The solid line shows the prescription by \citet{Kewley01}. Objects with AGN-like spectra are close to the upper right while star-forming galaxies reside closer to the lower left corner as labelled.}\label{fig:BPT}
\end{center}
\end{figure}

\begin{figure}
\begin{center}
\includegraphics[width=84mm]{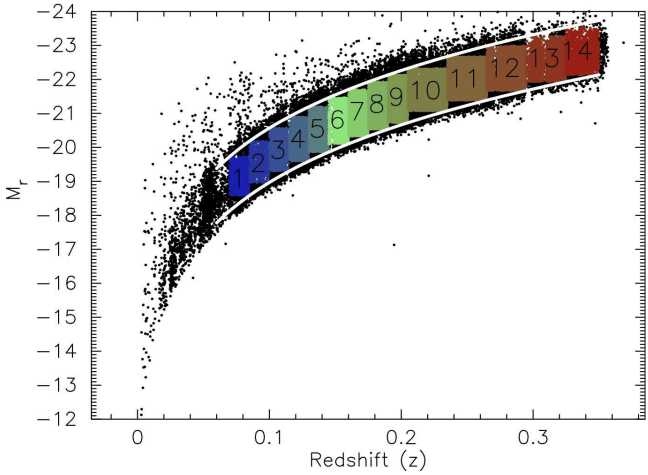}
\caption{The redshift and absolute $r$-band magnitude distribution of galaxies in our 14 GAMA volume limited samples (see Table \ref{table:samples}). White lines show the lower and upper apparent magnitude completeness of GAMA.}\label{fig:samples}
\end{center}
\end{figure}

\begin{table}
\begin{center}
\begin{tabular}{c c c c c}
\hline\hline
Sample&$z_{\rm min}$&$z_{\rm max}$&$M_{\rm r,max}$&$M_{\rm r,min}$\\
 (1) & (2) & (3) & (4) & (5)\\
\hline
1&0.070&0.085&-18.58&-19.73\\
2&0.085&0.100&-18.97&-20.18\\
3&0.100&0.115&-19.30&-20.57\\
4&0.115&0.130&-19.60&-20.90\\
5&0.130&0.145&-19.86&-21.20\\
6&0.145&0.160&-20.10&-21.46\\
7&0.160&0.175&-20.33&-21.70\\
8&0.175&0.190&-20.53&-21.93\\
9&0.190&0.205&-20.72&-22.13\\
10&0.205&0.235&-21.07&-22.32\\
11&0.235&0.265&-21.39&-22.67\\
12&0.265&0.295&-21.67&-22.99\\
13&0.295&0.325&-21.93&-23.27\\
14&0.325&0.350&-22.13&-23.53\\
\hline
\end{tabular}
\caption{Definition of the selected volume limited samples (column 1). Redshift limits are given in columns 2 and 3, while absolute Petrosian magnitude limits can be found in columns 4 and 5 for each sample.}
\label{table:samples}
\end{center}
\end{table}

\section{Data}\label{sec:data}
\subsection{GAMA survey}
Upon completion, the GAMA survey \citep[see][]{Driver11} will include data from many telescopes (AAT, VST, VISTA, {\em Herschel}, ASKAP, and GALEX) to create a database ranging from radio to ultraviolet wavelengths for roughly 400,000 galaxies over 360 square degrees. Optical spectra from AAOmega on the AAT have been obtained for $\sim$150,000 galaxies over three regions covering 144 square degrees in total as part of GAMA phase I. The GAMA Phase-I spectroscopic completeness is 98\% to an $r$-band Petrosian magnitude of 19.4 over two of the regions and $r\sim19.8$ over the third region. GAMA Phase-II will include a larger volume and will be complete to an $r$-band Petrosian magnitude of 19.8. Because of its depth, it is a good complement to SDSS probing to lower stellar masses and higher redshift. This work uses GAMA Phase-I data.

We provide a brief summary of the spectroscopic processing, with full details provided by Hopkins et al. (in preparation). The blue and red parts of the AAOmega optical spectra are processed and reduced separately using the {\sc 2dfdr} data reduction pipeline \citep[see][]{Croom04,Sharp10a} before combination. The pipeline performs standard bias subtraction, flat-fielding, tram-line fitting and wavelength calibration using arcs. Sky subtraction is done using dedicated sky fibres before performing optimal 1D extraction and is refined using Principal Component Analysis sky subtraction \citep{Sharp10b}. This is followed by flux calibration using standard star observations. Redshifts are derived from the 1D spectra following the process described in \citet{Driver11}.

\subsection{Emission line measurements}
We use the \emph{Gas AND Absorption Line Fitting algorithm} ({\sc GANDALF}, \citealt{Sarzi06}), an extension of the \emph{Penalized Pixel cross-correlation Fitting} ({\sc pPXF}) algorithm of \citet{Cappellari04}, to model the stellar absorption and emission line spectral components of our flux calibrated spectra. We measure the corrected flux and equivalent width of the strong emission lines from the GANDALF output emission line spectra. Our stellar population templates are from \citet{Maraston11} based on the code of \citet{Maraston05}, which uses the MILES stellar library \citep{SanchezBlazquez06} with the revised spectral resolution \citet{Beifiori11}. Our emission line measurements are effectively stellar-absorption corrected through the template fitting performed by GANDALF. We use the Balmer decrement to apply the dust obscuration curve correction of \citet{Cardelli89} to the emission line measurements, as recommended by \citet{Calzetti01}.

\subsection{Sample selection}\label{sec:sample}
We construct several subsamples of the GAMA data. First, we use the relationship from \citet{Kauffmann03} to discriminate galaxies with significant AGN contribution from star-forming galaxies using their position in the BPT diagram \citep{Baldwin81}. Fig.~\ref{fig:BPT} illustrates this selection. Galaxies whose ${\rm H\alpha}$ or ${\rm H\beta}$ lines are affected by overlapping strong sky line residuals are rejected. We nominally select galaxies for which the strong emission lines ${\rm H\alpha}$, ${\rm H\beta}$ and ${\rm [NII]\lambda6583}$ have $\rm S/N\ge3$. We refer to this selection as our ``fiducial'' selection throughout. A discussion of possible selection effects and justification of the fiducial selection are presented in Section \ref{sec:mmr}.  We reject any galaxies at $z<0.061$ to minimise fibre aperture issues as described in \citetalias{Kewley08} scaled for GAMA fibres (i.e., $2"$ in diameter). We also reject objects with a Balmer decrement $\rm H\alpha/H\beta<2.5$ and extinction $E(B-V)>10$. We refer to this sample as the `main sample' for our analysis.

Finally, we also select many volume limited samples in narrow redshift ranges, with maximum and minimum absolute $r$-band Petrosian magnitude determined based on the survey apparent magnitude completeness limits (i.e., $17.8\le r_{\rm petro}\le19.4$). This is done in order to have the broadest range in absolute magnitudes possible, yielding a maximal stellar mass range in each bin. The redshift ranges for each bin are given in Table~\ref{table:samples} and shown in Fig.~\ref{fig:samples}.

 \section{Analysis}\label{sec:analysis}
\subsection{Star formation rates}
We compute the SFR from the stellar absorption and extinction corrected H${\rm\alpha}$ flux measured by GANDALF on the flux calibrated spectra using equation~2 of \citet{Kennicutt98}. Because they are measured from the flux calibrated spectra, our SFR measurements include an implicit correction for fibre aperture effects. The implicit assumption made is that star formation is distributed across the galaxy following the stellar light \citep{Hopkins03}. Throughout this work, we assume a Salpeter IMF \citep{Salpeter55}. Our SFR measurements compare well with those published in \citet{Gunawardhana11} and SDSS-measured values in \citet{Brinchmann04}. Throughout, SFR is quoted in units of $\rm M_{\odot}\space yr^{-1}$.

\subsection{Stellar masses}
Stellar mass measurements are described in \citet{Taylor11}. They are determined using spectral energy distribution fitting of the $u$-, $g$-, $r$-, $i$- and $z$-magnitudes. We use a library of $\sim$80,000 templates from the \citet{BC03} stellar population models. The dust curve of \citet{Calzetti01} is assumed during this process. The quoted stellar mass (in units of $\rm M_{\odot}$) is derived from marginalisation over the posterior probability distribution of all fitted parameters (e.g., age, metallicity, stellar mass, star formation history, etc). An offset of 0.2 is applied to the log of the stellar mass values published in \citet{Taylor11} to accommodate our assumption of a Salpeter IMF.

\subsection{Gas-phase metallicities}
As stated in Section \ref{sec:introduction}, several methods may be used to derive gas-phase metallicities of star-forming galaxies \citep[see reviews by][]{LopezSanchez10,LopezSanchez12}. Although the direct method should be preferred, the faintness of the auroral lines (particularly, the $\rm[OIII]\lambda4363$ emission line) often prevents its use for relatively high metallicities (12+log(O/H)$\geq$8.4). Hence, techniques using the more readily measurable strong emission-lines have been developed to estimate galaxy gas-phase metallicity. In this work, we present a representative set of three of the popular strong emission line methods discussed in the literature. These include techniques based on photoionization models \citepalias{McGaugh91,Kewley02,Kobulnicky04} and empirical calibrations from samples of objects with known metallicity \citepalias{Pettini04}.

\subsubsection*{Method 1: \citet{Kewley02}}
First, we use the method described in \citetalias{Kewley02} with the update described in Appendix A of \citetalias{Kewley08}. This can be summarised as a combination of the $N2O2=\rm[N\thinspace\textsc{ii}]\lambda6583/[O\thinspace\textsc{ii}]\lambda3727$ ratio and an average of several calibrations of the degenerate $R23$ line ratio. For $N2O2>-1.2$ we solve the following polynomial:
\begin{eqnarray}
N2O2=1106.8660-532.15451Z+96.373260Z^2\\-7.8106123Z^3+0.23928247Z^4,\nonumber
\end{eqnarray}
where Z=log(O/H)+12. 

If $N2O2<-1.2$, we use the average of the \citet[][hereafter KK04]{Kobulnicky04} and \citet[][hereafter M91]{McGaugh91} methods. In M91, oxygen abundances ($Z_{\rm M91}$) for the lower R23 branch are derived using:
\begin{equation}
Z_{\rm M91}=12-4.944+0.767x+0.602x^2-y_1(0.29+0.332x-0.331x^2),
\end{equation}
where $y_1=\rm\log([O\thinspace\textsc{iii}]\lambda\lambda4959,5007/[O\thinspace\textsc{ii}]\lambda3727)$ and $x=\log R23$. \citetalias{Kobulnicky04} obtain oxygen abundances ($Z_{\rm KK04}$) by iterating the following two equations:
\begin{eqnarray}\label{eq:KK04q}
\log q=(32.81-1.153y_2^2+Z(-3.396-0.025y_2\\+0.1444y_2^2))\times(4.603-0.3119y_2-0.163y_2^2+Z\nonumber\\-0.48+0.0271y_2+0.02037y_2^2))^{-1}\nonumber,
\end{eqnarray}
where $y_2=\rm\log([O\thinspace\textsc{iii}]\lambda5007/[O\thinspace\textsc{ii}]\lambda3727)$ and
\begin{equation}\label{eq:KK04z}
Z_{\rm KK04}=9.40+4.65x-3.17x^2-\log q(0.272+0.547x-0.513x^2),
\end{equation}
with $q$ the ionization factor, until convergence is achieved. Hence, for $N2O2<-1.2$, the final quoted oxygen abundance is simply $12+\log(O/H)=(Z_{\rm M91}+Z_{\rm KK04})/2$.

\subsubsection*{Method 2: \citet{Pettini04}}
The second method used is that described in \citet[][hereafter PP04]{Pettini04}. Briefly, we use the $O3N2$ ratio, which is defined as follows:
\begin{equation}
O3N2=\log{\left(\rm{\frac{[OIII]\lambda5006/H\beta}{[NII]\lambda6583/H\alpha}}\right)}.
\end{equation}
In \citetalias{Pettini04}, the $O3N2$ ratio abundances are calibrated using HII regions yielding the following relationship:
\begin{equation}
{12 + \log (O/H)}_{PP04} = 8.73 - 0.32 \times O3N2.
\end{equation}
Because metallicities obtained directly using individual HII regions result in lower metallicities \citepalias[e.g.][]{Kewley02}, we use the public SDSS line measurement \citep{Brinchmann04} and oxygen abundances \citepalias{Tremonti04} to derive the following empirical conversion between \citetalias{Pettini04} and \citetalias{Tremonti04} abundances (see Lara-L\'opez et al., in preparation):
\begin{equation}\label{eq:PP04conv}
{12 + \log (O/H)}=0.103+1.021\times\left({\rm12 + log (O/H)}_{PP04}\right). 
\end{equation}
Equation \ref{eq:PP04conv} ``accommodates'' the well known 0.2-0.4 dex offset expected between metallicities derived using the direct and model based methods (see Section \ref{sec:introduction}).

\subsubsection*{Method 3: \citet{Kobulnicky04}}
The third abundance determination method presented in this work is that of \citetalias{Kobulnicky04} as updated by \citetalias{Kewley08}. It is based on the popular R23 ratio and hence yields representative results of studies that use this ratio. The lower R23 branch conversion is given by iterating Equations \ref{eq:KK04q} and \ref{eq:KK04z} until convergence. For $N2O2\ge-1.2$, metallicities are derived by iterating Equation \ref{eq:KK04q} and
\begin{eqnarray}
{12 + \log (O/H)}_{KK04}=9.72-0.777x-0.951x^2-0.072x^3\\-0.811x^4-\log(q)(0.0737-0.0713x-0.141x^2+0.0373x^3\nonumber\\-0.058x^4)\nonumber
\end{eqnarray}
until convergence.

\begin{figure*}
\begin{center}
\includegraphics[width=164mm]{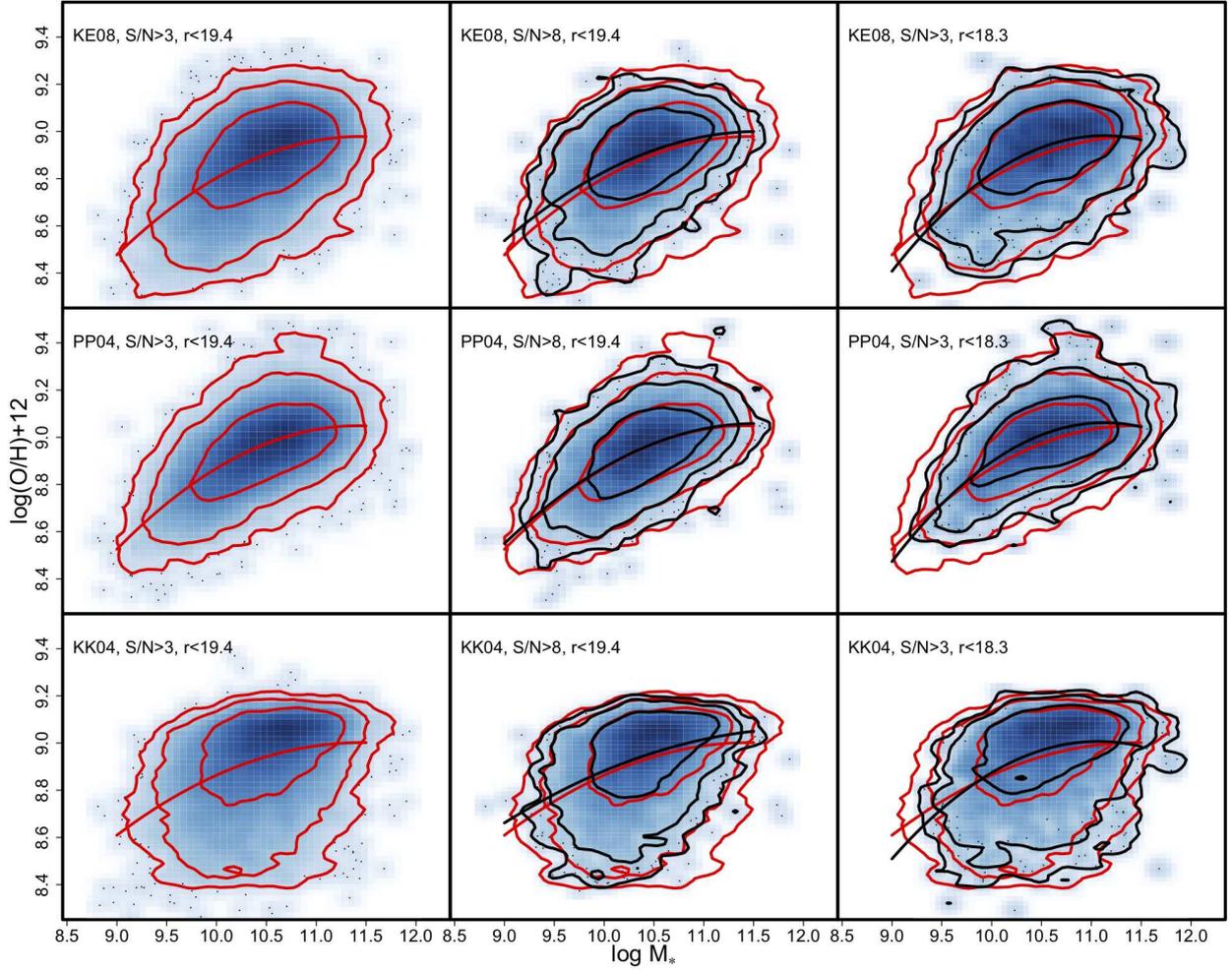}
\caption{A summary of the influence of varying the apparent selection criteria such as the survey apparent magnitude limit ($r$) and S/N limit in our fiducial lines ($\rm H\alpha$, $\rm H\beta$ and $\rm[NII]\lambda6583$) for three representative metallicity calibrations as labelled: \citetalias{Kewley08} (top row), \citetalias{Pettini04} (middle row) and \citetalias{Kobulnicky04} (lower row). For each row, red contours show the 1, 2 and 3 $\sigma$ profiles for our fiducial selection criteria, while the black contours show how varying S/N ratio cut and survey apparent magnitude limit influence the position and shape of the MMR. The red and black solid lines are respective second order polynomial fits to the MMR data. The fitted position of the MMR varies slightly for all ratios, but these variations are small (i.e., typically $<0.05$ dex).}\label{fig:MMRselsummary}
\end{center}
\end{figure*}

\subsubsection*{\bf Uncertainties}
In all cases, uncertainties are computed using Monte-Carlo methods. Assuming that the uncertainties on the measured emission line fluxes are normally distributed, we draw 1000 random measurements of the relevant emission lines for each object and recompute the metallicity each time. The quoted uncertainty for each calibration is the standard deviation of the relevant 1000 measurements.

For completeness, we also compute oxygen abundances using the methods described in \citetalias{Kewley02} for various available emission lines and the R23 calibrations of \citetalias{McGaugh91}, \citet{Zaritsky94} and \citetalias{Tremonti04}. We choose to present the above mentioned methods of \citetalias{Kewley08}, \citetalias{Pettini04} and the R23 calibration of \citetalias{Kobulnicky04} as a representative selection of the various methods used in the literature. We find that these three methods summarize and represent well the range of results obtained from the variety of measurement methods.

\section{Results}\label{sec:results}
This section focuses on the sensitivity of the fitted MMR with a set of common apparent selection criteria and abundance calibrations. We also look for evidence of redshift evolution and dependency on SFR. The three-dimensional distribution of galaxy stellar mass, metallicity and SFR in the joint GAMA and SDSS samples is discussed in detail in a companion paper (Lara-L\'opez et al., in preparation).

\begin{figure*}
\begin{center}
\includegraphics[width=165mm]{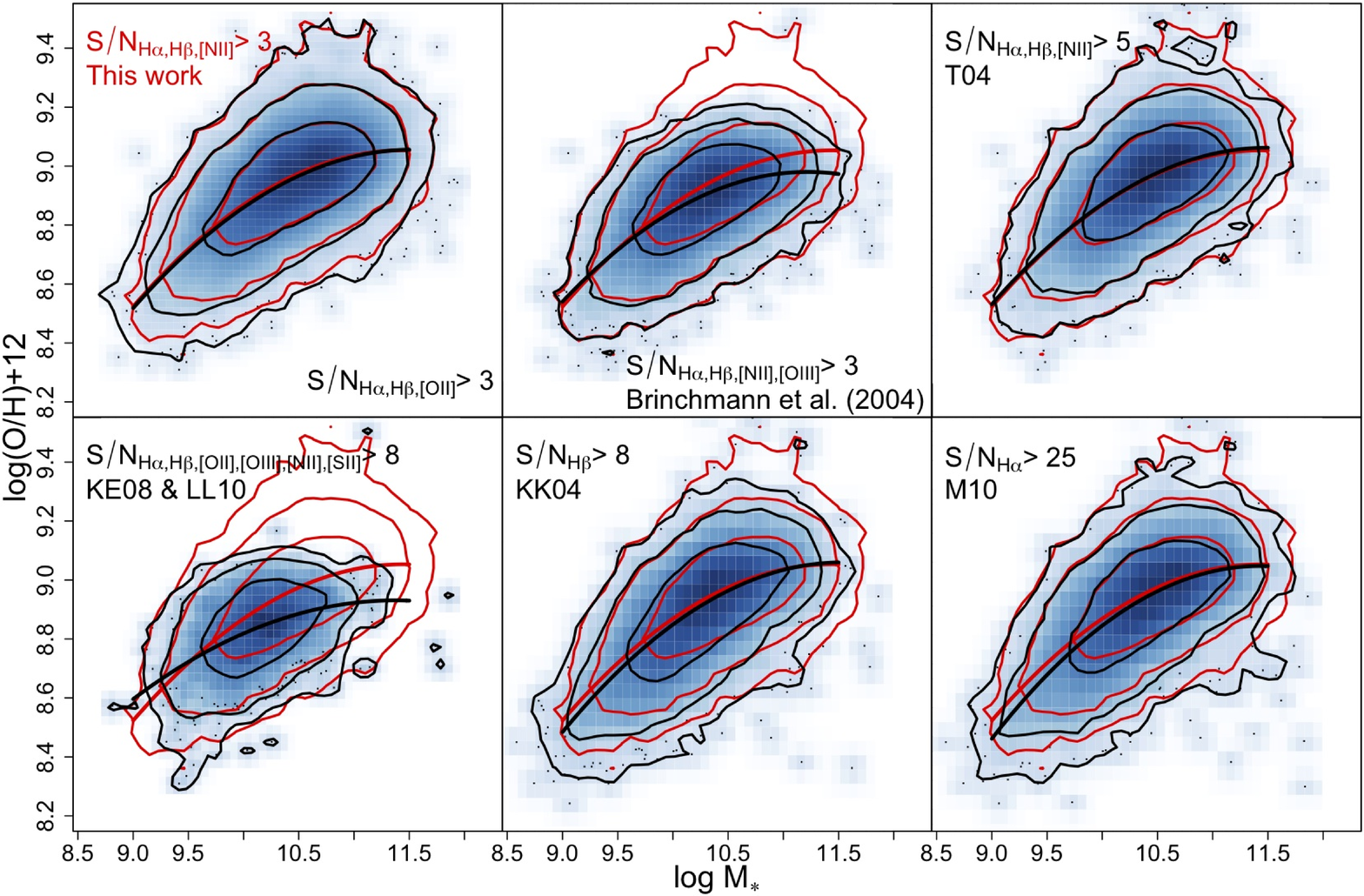}
\caption{GAMA MMR measured using the a variety of selection criteria taken mostly from the recent literature, as labeled, for the \citetalias{Pettini04} abundance estimate. The MMR obtained using the selection criteria of \citet{Brinchmann04}, \citetalias{Tremonti04}, \citetalias{Kewley08}, \citet[][LL10]{LaraLopez10a}, \citet[][KK04]{Kobulnicky04} and \citet[][M10]{Mannucci10} (respective black 1, 2 and 3$\sigma$ contours) are compared to that obtained using the selection criteria used in this work (red 1, 2 and 3$\sigma$ contours). The MMR measured varies significantly ($>0.05$ dex) if one selects on the $\rm[OIII]\lambda5007$ line.}\label{fig:MMRsamples}
\end{center}
\end{figure*}

\begin{figure*}
\begin{center}
\includegraphics[width=165mm]{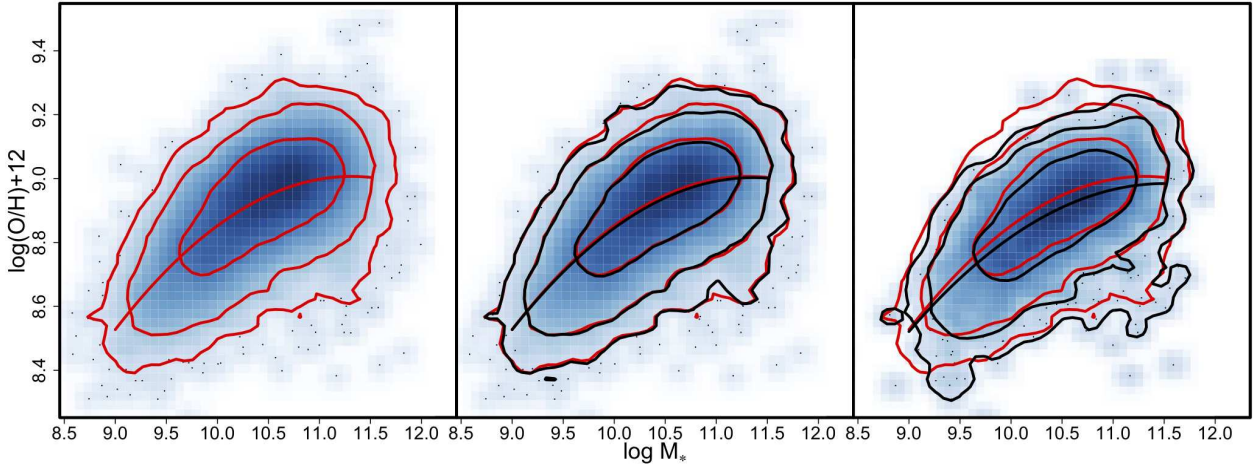}
\caption{GAMA MMR measured using galaxies selected with metallicity uncertainties less than 0.1 dex (left), 0.05 dex (middle) and 0.01 dex (right). Contours are similar to those of Fig. \ref{fig:MMRselsummary} and \ref{fig:MMRsamples}.}\label{fig:errors}
\end{center}
\end{figure*}

\begin{figure*}
\begin{center}
\includegraphics[width=70mm]{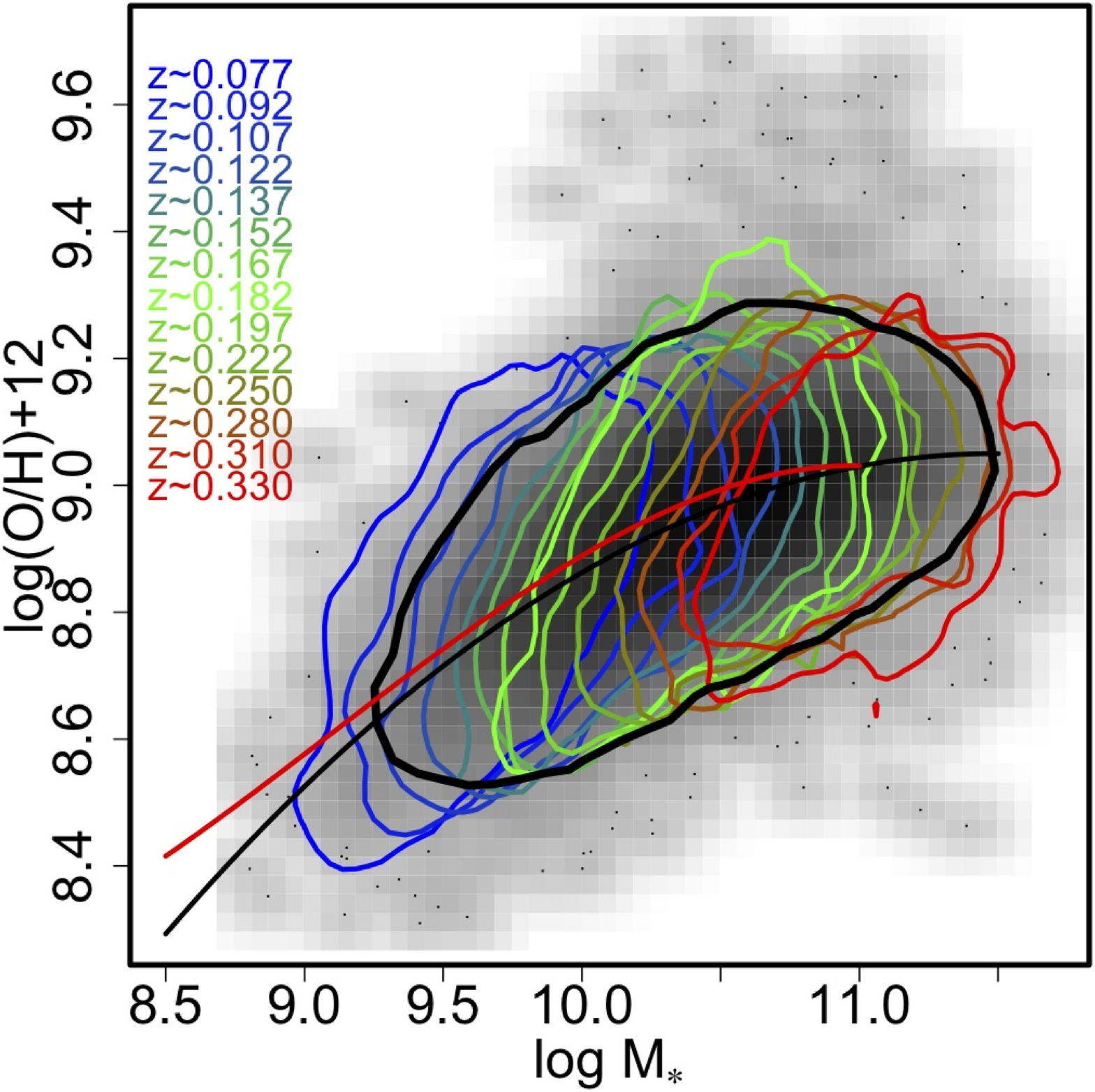}\includegraphics[width=70mm]{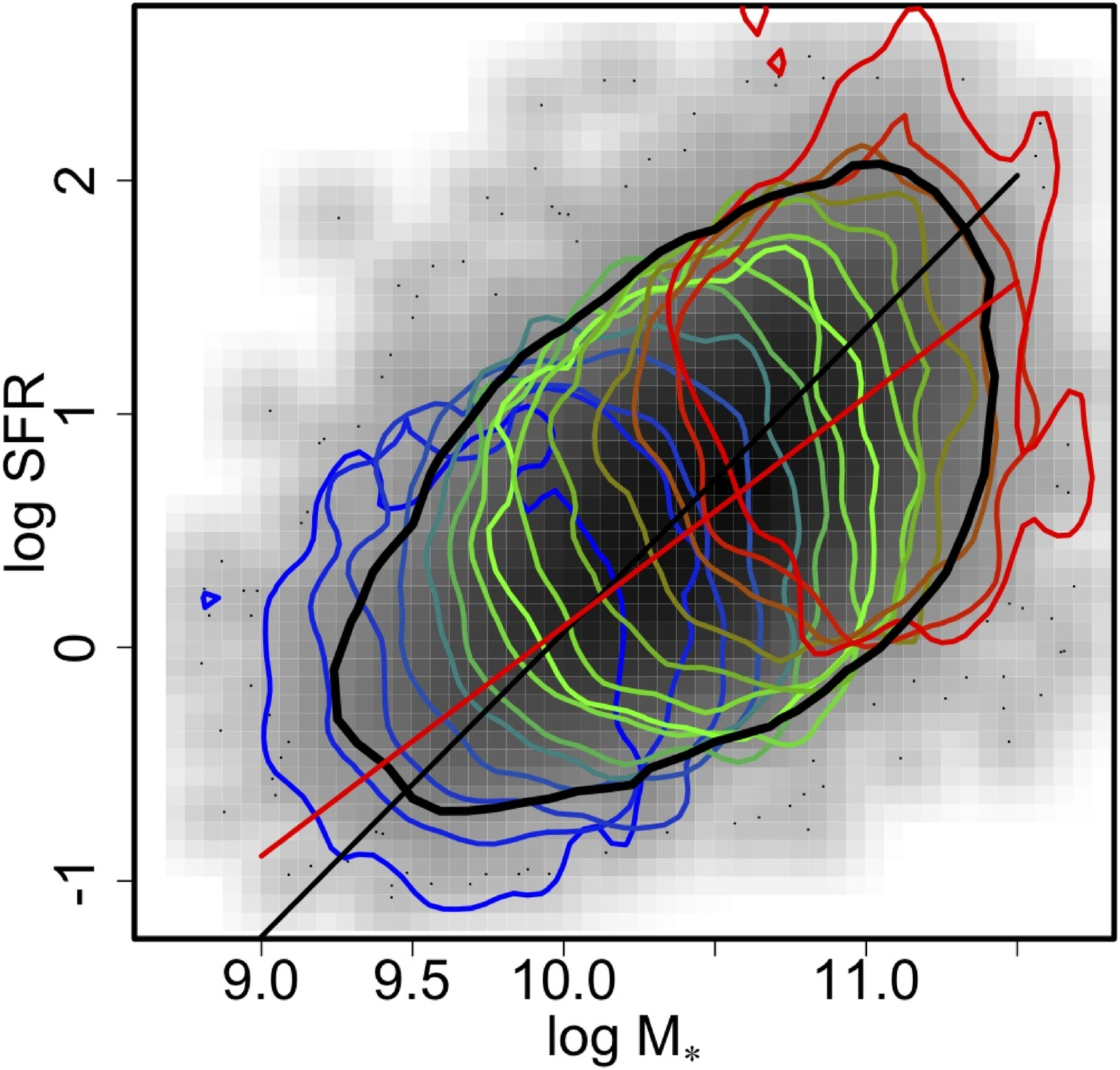}\\
\includegraphics[width=70mm]{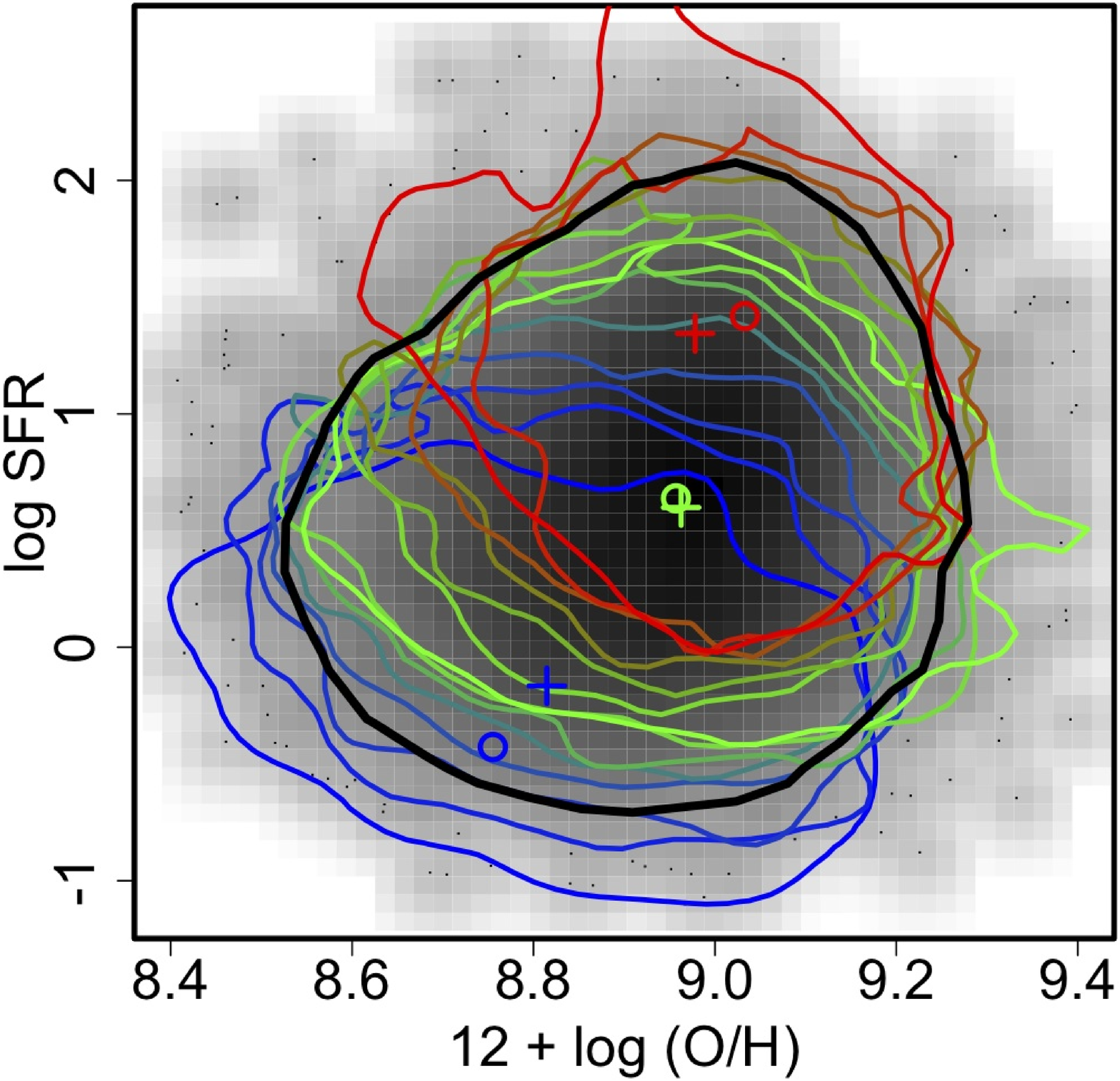}
\caption{The GAMA mass-metallicity (upper left), mass-SFR(upper right) and metallicity-SFR (lower-middle) relationships in our main sample. Black and coloured contours show 2$\rm\sigma$ contours for the main and volume limited samples with median redshifts as labelled, respectively. Red and black solid lines show the fitted SDSS MMR from \citetalias{Kewley08} using the updated \citetalias{Kewley02} conversion and our fit to the GAMA data, respectively. Open circles in the lower panel show the expected mean metallicity and SFR as inferred from the mean mass using Equations~\ref{eq:gamammr} and \ref{eq:MSFR} for the lowest, intermediate and highest redshift bins. The actual mean metallicity and SFR for these samples are shown with crosses (see Section \ref{sec:ZSFR}).}\label{fig:MZSFR-vol}
\end{center}
\end{figure*}

\subsection{Mass-metallicity relation}\label{sec:mmr}
As mentioned in Section~\ref{sec:sample}, we study the dependence of the MMR on selection criteria such as the signal-to-noise (S/N) in various lines and apparent magnitude completeness, as well as on the metallicity calibration used. For this, we compare the MMR obtained with our fiducial selection criteria with that obtained by increasing the S/N limit from 3 to 8, but keeping the $r$-band Petrosian magnitude limit constant. Next, we vary the $r$-band Petrosian magnitude limit between 19.4 and 18.3, while keeping the S/N limit at 3. In all cases, we fit a second order polynomial to the distribution obtained, following the approach of other recent work. In what follows, we consider vertical (metallicity) variations of order $0.05$ dex to be significant, as they could account for at least half of the reported variations of the MMR with redshift for $z\lesssim0.4$ \citep[see for example][]{LaraLopez09b} and specific SFR \citep[e.g.,][]{Ellison08}. In other words, we are looking for vertical differences of either the contours or fitted MMR with variations of the selection. Fig.~\ref{fig:MMRselsummary} summarizes the results of this exercise. Varying these selection criteria for the MMR does not significantly alter the fitted MMR for a given calibration, however as emphasised by \citetalias{Kewley08}, the shape and location of the MMR is significantly altered between the three calibrations shown. 

Fig.~\ref{fig:MMRsamples} shows that the MMR can be significantly different depending on the choice of selection criteria, for a variety of criteria that have been used in recent work. In general, we find that selecting on the $\rm[OIII]\lambda5007$ line most influences the position and shape of the measured MMR. This is a consequence of higher metallicities being associated with relatively fainter oxygen lines, which are more likely to show lower S/N. We note that the effect is more significant with the higher S/N threshold applied by KE08 and LL10, compared to the lower threshold applied by \citet{Brinchmann04}, although even this cut introduces a measurable effect. We therefore recommend caution, and rigorously consistent sample selection if cuts are being made based on the S/N ratio in these lines, when studying small variations of the MMR.

An alternative way to identify galaxies with reliable metallicity measurements is to select based on the size of the metallicity uncertainty itself. The effects of such a selection are shown in Fig. \ref{fig:errors}, where we vary the selection on metallicity uncertainty from 0.1 to 0.01 dex. It can be seen that for reasonably moderate criteria, the MMR and contours are essentially unaltered, however applying an extreme cut does alter the shape and position of the MMR. We cannot reject the possibility that these could be due to genuine evolutionary effects since galaxies at lower redshift would inevitably have lower errors. More importantly, we insist that \emph{selecting based on overall metallicity uncertainties can only be applied for calibrations that are monotonic and smooth conversions between observables and metallicity}. Piecewise calibrations such as the \citetalias{Kewley02} and R23-based methods have discontinuous conversion functions that yield large errors around the transition, hence a selection on error would discriminate against galaxies of those or similar metallicities (typically $12+\log(O/H)\sim$ 8.4 dex).

Fig.~\ref{fig:MZSFR-vol} shows the MMR, the mass-SFR relation and the metallicity-SFR relation for GAMA. Each panel shows the relation for our main sample in greyscale, with the distribution for each of the 14 volume limited redshift samples overlayed as coloured contours. We emphasize that even though our volume limited samples are complete in a luminosity sense (absolute r-band magnitude), this does not correspond to a strict stellar mass limit. While each volume limited sample is itself complete, we observe the expected Malmquist-type bias as we consider samples selected at different redshift. Galaxies of low (high) stellar masses are preferentially observed in the lowest (highest) redshift samples. As a result, in each redshift bin, the range of stellar masses probed is limited. This complicates the study of possible evolution within GAMA. This limitation is addressed by combining the SDSS and GAMA samples by Lara-L\'opez et al. (in preparation).

Bearing all previously mentioned caveats in mind, we tentatively find that the MMR in GAMA is mostly consistent with that measured in SDSS. It is important to emphasize that even though GAMA metallicities are scaled to the same calibration as that used in the presented SDSS MMR they are not strictly measured using the same selection criteria and metallicity diagnostic. Hence, for the reasons already outlined above, its exact position and shape is not directly comparable to the GAMA MMR presented here. Nevertheless, the similarity between the SDSS and GAMA MMRs is encouraging to see given the inherently different apparent magnitude ranges probed by the two surveys. Indeed, although GAMA covers a smaller volume than SDSS, it is probing to fainter magnitudes, and a galaxy of a given mass will on average be detected at higher redshift in GAMA than in SDSS. This consideration suggests that any evolutionary effects over this redshift range are likely to be small. A companion paper (Lara-L\'opez et al., in preparation) compares the SDSS and GAMA surveys directly and self-consistently.

%

As is common, we fit a second order polynomial to the MMR in GAMA and obtain 
\begin{eqnarray}\label{eq:gamammr}
\log(O/H)+12=(-2.2\pm0.3)+(1.95\pm0.07)\times\log(M_*)\\-(0.085\pm0.003)\times\log(M_*)^2\nonumber
\end{eqnarray}
with an $rms$ scatter of $\sigma=0.12$ dex, similar to that found for SDSS \citep[e.g.,][]{Tremonti04,Kewley08}. Equation \ref{eq:gamammr} is valid for $8.7 \le \log(M_* (M_\odot))\le12.5$ for our fiducial selection criteria and the \citetalias{Pettini04} metallicity calibration. We emphasize here that the MMR fit in this fashion arises from the contribution of galaxies at all redshifts probed in GAMA ($0.061\le z \lesssim 0.4$). As a result, galaxies at low (high) redshift generally contribute to the low (high) mass end of the MMR, and hence possible evolutionary processes are likely to be masked.

\subsection{Mass-SFR relation}\label{sec:MSFR}
The upper right panel of Fig. \ref{fig:MZSFR-vol} shows the relationship between stellar mass and SFR in GAMA for our main sample as well as the 14 volume limited redshift samples. While the relationship is difficult to detect in the individual volume limited samples due to the small stellar mass range probed in each one, we find a positive correlation between $\log(M_*)$ and SFR in the main sample, spanning the full redshift range probed. We use a bisector fit for galaxies in the main sample with reliable stellar mass and SFR measurements after applying a 2$\sigma$ clipping. We obtain:
\begin{equation}\label{eq:MSFR} 
\log(SFR)=(1.304\pm0.008)\times(\log(M_*))-(12.98\pm0.08)
\end{equation}
with a normal $rms$ scatter of $\sigma\sim 0.5$ dex. As with the MMR, we find that all the volume-limited samples are consistent with the global fit, and the overall trend is consistent with that found in the SDSS, although the global fit is still affected by the Malmquist bias.

\subsection{Metallicity-SFR relation}\label{sec:ZSFR}
 
The lower panel in Fig.~\ref{fig:MZSFR-vol} shows the relationship between metallicity and SFR for galaxies in GAMA. We find only a weak relation between the two in only some of our volume-limited redshift samples. The main sample itself doesn't exhibit any correlation. In order to better understand the effect of the MMR and mass-SFR relationships measured in Sections~\ref{sec:mmr} and \ref{sec:MSFR} on the metallicity-SFR distribution, we perform the following exercise. We compute the mean stellar mass in each volume-limited sample and use Equations~\ref{eq:gamammr} and \ref{eq:MSFR} to infer a corresponding metallicity and SFR, respectively. These are shown for the lowest, intermediate and highest redshift bins as open circles in the lower panel of Fig.~\ref{fig:MZSFR-vol}. The actual mean metallicity and SFR for these samples are shown with crosses. While Equations~\ref{eq:gamammr} and \ref{eq:MSFR} provide a consistent estimate for the position of the data at the intermediate redshift samples, they seem to over- (under-)predict the metallicity at low (high) redshift by $\sim0.05$ dex. This result is likely to arise as a consequence of the fits to the whole sample being dominated by the intermediate redshift samples where the data density is greatest. The implication, though, is that at the higher-redshift end the observed metallicity is slightly lower than would be consistent with the extrapolation from the intermediate redshift objects. Conversely, at the lower-redshift end, the observed metallicity (and SFR) is slightly higher ($\sim0.2$ dex). This may be suggestive of evolutionary effects, even over this redshift range (see also Lara-L\o'pez et al., in preparation).

\section{Summary and Conclusions}\label{sec:concl}

We have presented the dependence of the mass-metallicity relationship on a variety of selection criteria, metallicity calibrations and SFR. As in previous studies, we see the strong mass-metallicity relationship for the GAMA samples. The shape and position of the mass-metallicity relationship can vary significantly depending on the metallicity calibration and selection used. In particular, we find that selecting on the $\rm[OIII]\lambda5007$ line most affects the position and shape of the mass-metallicity relationship, while varying survey depth does not. Alternatively, for monotonic metalllicity calibrations such as that of \citet{Pettini04}, one can also select on the metallicity uncertainties itself without biasing the sample. Given these, caution must be used when comparing the mass-metallicity relationship obtained using different metallicity calibrations and/or sample selection criteria.

The mass-metallicity relationship in GAMA is in reasonable agreement with that found in SDSS despite the very different stellar mass range probed at any given redshift. Notwithstanding the cautions illustrated in this work, this result implies that evolutionary effects out to $z<0.35$ are likely to be small. Such small effects reinforce the importance of caution in selecting samples with which to estimate their measurement.

Using volume-limited redshift samples, we find tentative evidence that galaxies at lower redshift may indeed have measurably higher metallicities than those at higher redshift, even over the redshift range probed within GAMA. This suggests that there could be detectable evolution of the MMR on the order of a few tenths of a dex in metallicity spanning $0.06<z<0.35$.

In a companion paper (Lara-L\'opez et al., in preparation), we investigate the joint SDSS + GAMA mass-metallicity relationship, its evolution and the joint 3D distribution in metallicity-$\log(M_*)$-SFR space self-consistently and in detail.

Finally, while the use of fibre-fed spectrographs has enabled large redshift surveys up to now, all are essentially limited by aperture biases. Indeed, it is imperative to test the basic assumptions generally used in similar work that 1) star formation follows the distribution of stellar light and 2) models of single HII regions can be applied to spectra of entire galaxies. These assumptions can now be tested using available integral field unit spectrographs by observing galaxies at low redshift. Moreover, the Sydney-AAO Multi-object Integral field spectrograph with its hexabundle technology \citet{Croom12} will allow large spatially resolved spectroscopic galaxy surveys, enabling us to test these important assumptions and eliminating aperture biases.

\begin{acknowledgements}
CF thanks the Australian Astronomical Observatory for financial support in the form of a graduate top-up scholarship. This work was co-funded under the Marie Curie Actions of the European Commission (FP7-COFUND). GAMA is a joint European-Australasian project based around a spectroscopic campaign using the Anglo-Australian Telescope. The GAMA input catalogue is based on data taken from the Sloan Digital Sky Survey and the UKIRT Infrared Deep Sky Survey. Complementary imaging of the GAMA regions is being obtained by a number of independent survey programs including GALEX MIS, VST KIDS, VISTA VIKING, WISE, Herschel-ATLAS, GMRT and ASKAP providing UV to radio coverage. GAMA is funded by the STFC (UK), the ARC (Australia), the AAO, and the participating institutions. The GAMA website is http://www.gama-survey.org/ .
\end{acknowledgements}

\bibliographystyle{aa}
\bibliography{biblio}

\end{document}